\documentclass[
    ,final            
  ]
  {aipproc}

\layoutstyle{6x9}

\begin{document}

\title[$\varepsilon$ Indi Ba, Bb: a spectroscopic study]{$\varepsilon$ Indi Ba, Bb: a spectroscopic study of the nearest
known brown dwarfs}


\author{Robert R. King}{
  address={School of Physics, University of Exeter, Stocker Road, Exeter EX4 4QL,
UK}
}

\author{Mark J. McCaughrean}{
  address={School of Physics, University of Exeter, Stocker Road, Exeter EX4 4QL, UK}
}

\author{Derek Homeier}{
  address={Institut f\"{u}r Astrophysik, Geoerg-August-Universit\"{a}t,
\\Friedrich-Hund-Platz 1, 3077 G\"{o}ttingen, Germany}
}

\author{France Allard}{
  address={Centre de Recherche Astrophysique de Lyon, UMR 5574: CNRS, Universit\'{e} de
Lyon \'{E}cole Normale Sup\'{e}rieure de Lyon, 46 all\'{e}e d'Italie, 69364 Lyon Cedex
07, France}
}

\author{Ralf-Dieter Scholz}{
  address={Astrophysikalisches Institut Potsdam, An der Sternwarte 16, 14482 Potsdam,
Germany}
}

\author{Nicolas Lodieu}{
  address={Instituto de Astrof\'{i}sica de Canarias, V\'{i}a \'{L}actea s/n, E-38205 La
Laguna, Tenerife, Spain}
}

\begin{abstract}

The discovery of $\varepsilon$ Indi Ba and Bb, a nearby binary brown dwarf system with a
main-sequence companion, allows a concerted campaign to characterise the physical
parameters of two T dwarfs providing benchmarks against which atmospheric and
evolutionary models can be tested.  Some recent observations suggest the models at low
mass and intermediate age may not reflect reality with, however, few conclusive tests.

We are carrying out a comprehensive characterisation of these, the nearest known brown
dwarfs, to allow constraints to be placed upon models of cool field dwarfs. We present
broadband photometry from the $V$- to $M'$-band and the individual spectrum of both
components from 0.6-5.1$\mu$m at a resolution of up to R$\sim$5000. A custom analytic
profile fitting routine was implemented to extract the blended spectra and photometry of
both components separated by $\sim$0.7". We confirm the spectral types to be T1 and T6,
and notably, we do not detect lithium at 6708\AA~in the more massive object which may be
indicative both of the age of the system and the mass of the components.

\end{abstract}

\maketitle


\section{Introduction}

Studies of low mass stars and brown dwarfs are hampered by a lack of fundamental
measurements: radii, masses, ages, and high resolution atmospheric observations. This is
due to a number of factors including their intrinsic faintness, relatively low numbers of
known objects, and the mass-age degeneracy of sub-stellar objects. Binary systems have an
important role to play here: they can allow the determination of dynamical masses and
radii, they provide the laboratory in which to compare objects at the same age and
chemical composition, and where the systems contain previously well-studied main-sequence
stars, they provide external determinations of metallicity and age which field objects
lack, thus breaking the sub-stellar mass-age degeneracy. Previous studies which have
attempted to constrain low mass evolutionary models have been hampered by ambiguous ages,
possible unresolved binarity, and the difficulty of acquiring observations of close,
faint companions (cf. AB Dor C; \citet{Close:2005}, \citet{Luhman:2006}). Spectroscopic
observations of T dwarfs have mostly been either low resolution studies which allow
spectral classification and overall SED modelling to determine luminosities, or high
resolution studies of relatively small wavelength regions, for example, to investigate
gravity and effective temperature sensitive features.  Here we present high
signal-to-noise, moderate resolution optical to thermal-IR spectra of two T dwarfs which
have important external constraints from their parent star, $\varepsilon$ Indi A.  A
detailed analysis and comparison to atmospheric models is underway (King et al. 2008, in
prep.).

\section{Association with $\varepsilon$ Indi A}

The discovery of a very wide companion ($\sim$1500AU) to the high proper-motion
($\sim$4.7"/yr) $\varepsilon$ Indi was reported by \citet{Scholz:2003} in 2003. Being
amongst the closest known stars to our Solar System, $\varepsilon$ Indi A has a
well-constrained parallax from Hipparcos putting the system at a distance of
3.626$\pm$0.009pc.  This initial discovery was followed by the discovery of the
companion's binary nature by \citet{McCaughrean:2004}.  Model comparisons placed the
masses at 47$\pm$10 and 28$\pm$7 M$_{\rm{Jup}}$ for an age of 1.3Gyr. 

$\varepsilon$ Indi Ba and Bb are uniquely well-suited to provide key insights into the
physics, chemistry, and evolution of sub-stellar sources. Although there are a number of
other T dwarfs in binary systems, $\varepsilon$ Indi B has a very well-determined
distance, a reasonably well-determined age and metallicity, and has a short enough orbit
($\sim$13 years) such that the system and individual dynamical masses can soon be
determined (McCaughrean et al. 2008, in prep.; Cardoso et al., these proceedings).  It is
also bright and sufficiently separated to allow detailed photometric and spectroscopic
studies of both components.  Characterisation of this system will allow the
mass-luminosity relation at low masses and intermediate age to be tested; will allow
investigation of the atmospheric chemistry, such as the chemical equilibrium of CO and
CH$_4$, and will allow detailed investigation of the species in the atmosphere whilst
matching the overall spectral morphology caused by very broad absorption lines.

\section{Observations}

$\varepsilon$ Indi Ba, Bb was observed with the ESO VLT using FORS2 for optical
photometry and spectroscopy, and ISAAC for near- and thermal-IR photometry and
spectroscopy. In both the optical and near-IR observations the point-spread functions
(PSFs) of the two sources were blended due to the $\sim$0.7" separation of the binary
with typical 0.6" seeing. 

\begin{figure}
  \includegraphics[height=.25\textheight]{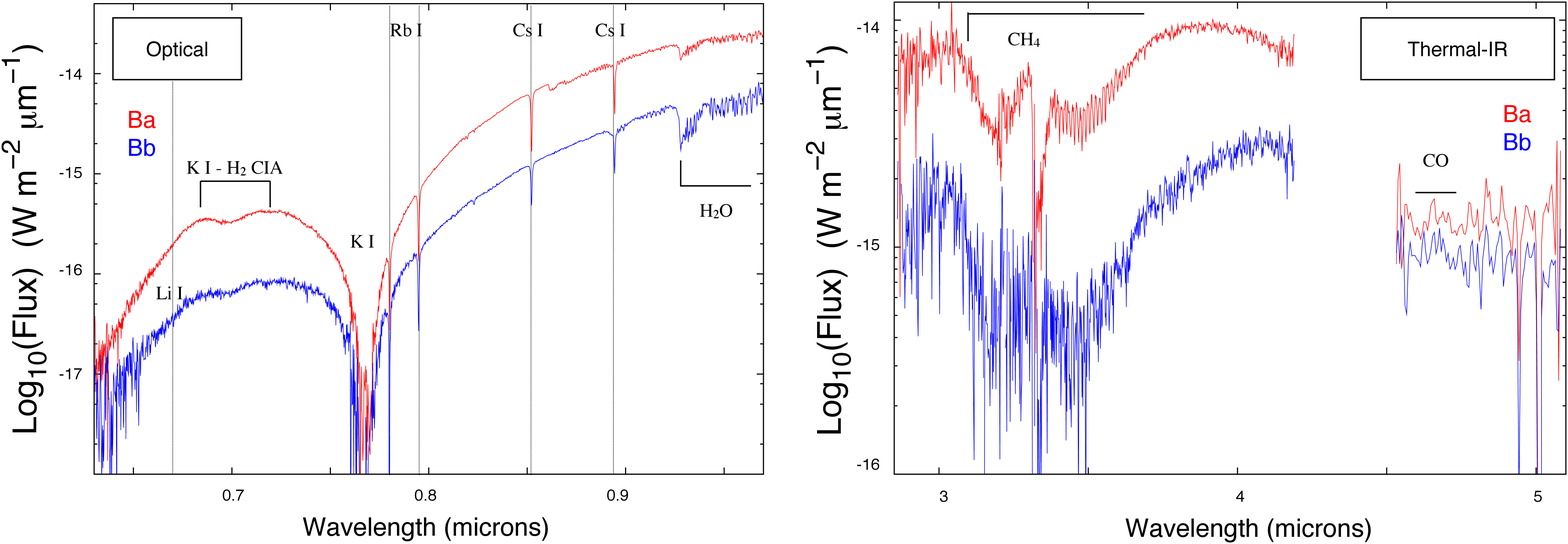}

  \caption{The optical and thermal-IR spectra of $\varepsilon$ Indi Ba (red, upper line)
and Bb. Note the lack of obvious lithium absorption at 6708\AA~(marked) in the brighter
component. The two deep absorption features near 5$\mu$m are residuals of bright telluric
emission.}

  \label{fig:a}
\end{figure}

\begin{figure}
  \includegraphics[width=\columnwidth]{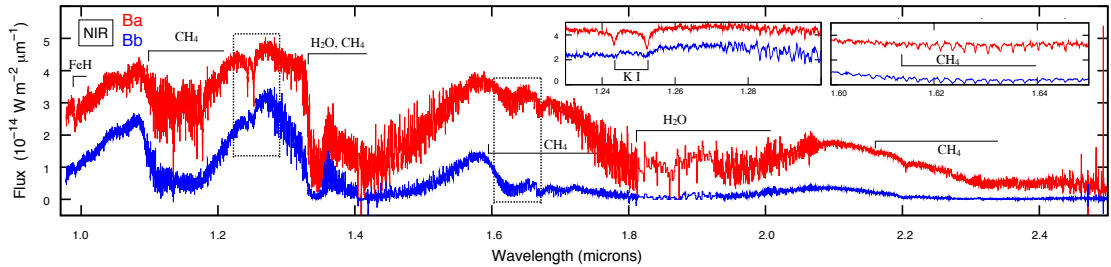}

  \caption{The near-IR spectra of $\varepsilon$ Indi Ba (red, upper line) and Bb. The
regions marked with dotted rectangles are shown inset, note the y-dispersion is
indicative of the wealth of real features: this is not noise.}

  \label{fig:b}
\end{figure}

The individual sources were extracted using a custom analytic profile fitting routine
applied to both the imaging and spectroscopy.   The results from these fits were
confirmed by extracting the optical photometry with IRAF/DAOPHOT as these were the only
observations with sufficient neighbours with which to model the PSF.  The derived
magnitudes are listed in Table \ref{tab:a} and the optical to thermal-IR spectra are
shown in Figures \ref{fig:a} and \ref{fig:b}.  The optical spectrum of $\varepsilon$ Indi
Ba lacks lithium absorption at 6708\AA~which may be expected in spectra at this
signal-to-noise and resolution for such an object below the lithium burning mass,
indicating a mass in excess of the previously derived model mass of 47 M$_{\rm{Jup}}$.
This is consistent with our revised age estimate of $\sim$5 Gyr for $\varepsilon$ Indi A
which will be discussed in a forthcoming paper (King et al. 2008, in prep.).

\subsection{Optical, near-IR, and thermal-IR spectra}

The proximity and separation of this system allow these moderate resolution spectra to be
acquired for both objects over a wide wavelength range which is practical for few other T
dwarfs. Here we show the full resolution spectra of both T dwarfs from 0.63-5.1$\mu$m
(Figs. \ref{fig:a} and \ref{fig:b}). The optical has a maximum S/N$\sim$350
(R$\sim$1000-2000), while the 1-2.5$\mu$m spectra have S/N$\sim$50-100 (R$\sim$5000) at
the $JHK$-band peaks. The $L$- and $M'$-band spectra have S/N$\sim$5-40 (R$\sim$500 and
250, respectively).  These can place important constraints on the atmospheric chemistry
of T dwarfs. The 1.81-1.92$\mu$m regions have been median filtered to highlight the
measure of the continuum level in this region of high telluric absorption. In particular,
the vertical dispersion seen in the $JHK$-band spectra is not noise - these are real
features. The dotted regions are shown in the inset plots.  Notice the features present
in both objects, some of which are stronger in $\varepsilon$ Indi Bb.

\section{Conclusions}

We have presented an overview of our comprehensive spectroscopic study of the individual
components of the nearest known binary brown dwarf system, $\varepsilon$ Indi Ba, Bb. 
The relative proximity of these T1 and T6 dwarfs ensures very high quality data and the
parent star, $\varepsilon$ Indi A, provides invaluable external constraints. In
particular, $\varepsilon$ Indi Ba shows no Li6708\AA~absorption, in contrast with some
model predictions for younger, lower-mass sources. This absence of lithium is consistent
with revised, higher age estimates from $\varepsilon$ Indi A. Our spectrum of
$\varepsilon$ Indi Bb is consistent with either a non-detection or minimal
Li6708\AA~absorption which would be expected at these lower temperatures where most of
the monatomic lithium has been sequestered into LiCl and other molecules.

We plan to use VLT NACO and FORS2 astrometry data to search for any evidence of medium-
to long-term photometric variability in these two T dwarfs. This may be complemented by a
VLT/X-Shooter proposal to search for possible spectral variability signatures and to
acquire a luminosity measurement at a single epoch.


\begin{table}
\begin{tabular}{lrr}
\hline
\hline
  \tablehead{1}{r}{b}{Filter}  
  & \tablehead{1}{r}{b}{$\varepsilon$ Indi Ba}
  & \tablehead{1}{r}{b}{$\varepsilon$ Indi Bb}   \\
\hline
$V$ & 24.47$\pm$0.04 & 26.92$\pm$0.09\\
$R$ & 20.08$\pm$0.02 & 21.10$\pm$0.05\\
$I$ & 16.78$\pm$0.03 & 18.51$\pm$0.03\\
$z$ & 15.20$\pm$0.03 & 16.57$\pm$0.04\\
$J$ & 12.20$\pm$0.03 & 12.96$\pm$0.03\\
$H$ & 11.60$\pm$0.02 & 13.40$\pm$0.03\\
$K$ & 11.42$\pm$0.02 & 13.64$\pm$0.02\\
$L'$ & 9.71$\pm$0.05 & 11.33$\pm$0.06\\
$M'$ & 10.67$\pm$0.23 & 11.04$\pm$0.23\\
\hline
\end{tabular}
\caption{Photometry of $\varepsilon$ Indi Ba and Bb}
\label{tab:a}
\end{table}


\begin{theacknowledgments}

RRK acknowledges the support of an STFC studentship.  Part of this work was funded by the
EC MC RTN CONSTELLATION (MRTN-CT-2006-035890)

\end{theacknowledgments}



\bibliographystyle{aipproc}   

\bibliography{ms}

\begin{thebibliography}{4}
\expandafter\ifx\csname natexlab\endcsname\relax\def\natexlab#1{#1}\fi
\providecommand{\enquote}[1]{``#1''}
\expandafter\ifx\csname url\endcsname\relax
  \def\url#1{\texttt{#1}}\fi
\expandafter\ifx\csname urlprefix\endcsname\relax\def\urlprefix{URL }\fi
\providecommand{\eprint}[2][]{\url{#2}}

\bibitem[{Close} et~al.(2005)]{Close:2005}
L.~M. {Close}, R.~{Lenzen}, J.~C. {Guirado}, E.~L. {Nielsen}, E.~E. {Mamajek},
  W.~{Brandner}, M.~{Hartung}, C.~{Lidman}, and B.~{Biller}, \emph{Nature}
  \textbf{433}, 286--289 (2005).

\bibitem[{Luhman} and {Potter}(2006)]{Luhman:2006}
K.~L. {Luhman}, and D.~{Potter}, \emph{Astrophysical Journal} \textbf{638},
  887--896 (2006).

\bibitem[{Scholz} et~al.(2003)]{Scholz:2003}
R.-D. {Scholz}, M.~J. {McCaughrean}, N.~{Lodieu}, and B.~{Kuhlbrodt},
  \emph{Astronomy \& Astrophysics} \textbf{398}, L29--L33 (2003).

\bibitem[{McCaughrean} et~al.(2004)]{McCaughrean:2004}
M.~J. {McCaughrean}, L.~M. {Close}, R.-D. {Scholz}, R.~{Lenzen}, B.~{Biller},
  W.~{Brandner}, M.~{Hartung}, and N.~{Lodieu}, \emph{Astronomy \&
  Astrophysics} \textbf{413}, 1029--1036 (2004).

\end{thebibliography}



\end{document}